\documentstyle[12pt]{article}
\input psfig
\def\be{\begin{equation}}
\def\ee{\end{equation}}
\def\ba{\begin{array}}
\def\ea{\end{array}}
\def\beqn{\begin{eqnarray}}
\def\eeqn{\end{eqnarray}}
\def\bt{\begin{tabular}}
\def\et{\end{tabular}}
\def\bc{\begin{center}}
\def\ec{\end{center}}

\def\vud{$|V_{ud}|$}
\def\vus{$|V_{us}|$}
\def\vcb{$|V_{cb}|$}

\def\vcd{$|V_{cd}|$}
\def\vcs{$|V_{cs}|$}

\def\vckm{$|V_{CKM}|$}
\def\rub{$|\frac {V_{ub}}{V_{cb}}|$}
\def\mu{$m_u$}

\begin{document}
\title{Implications of the Unitarity Triangle `uc' for J, $\delta$
and \vckm ~elements. }
\author{Monika Randhawa$^1$, V. Bhatnagar$^1$, P.S. Gill$^{1,2}$ and
 M. Gupta$^1$ \\
$^1$  {\it Department of Physics, Panjab University, Chandigarh-
  160 014, India.}\\
 $^2$  {\it Sri Guru Gobind Singh College, Chandigarh-160 026, India.}}
%  \date{11 March 1999}
 \maketitle

\begin{abstract}
  The Jarlskog rephasing invariant parameter $|J|$ is evaluated using
  one of the six Unitarity Triangles involving  well known CKM
  matrix elements \vud, \vus,~\rub, ~\vcd, ~\vcs~ and ~\vcb.
  With  PDG2000 values of \vud~ etc. as input, 
  we obtain $|J|=(2.71 \pm  1.12)  \times 10^{-5}$, which in 
 the PDG representation
 of CKM matrix leads to the  range  $21^o~to~159^o$
   for the CP violating phase $\delta$.
The CKM matrix elements evaluated using this range of $\delta$
  are in agreement with the PDG CKM matrix.
  The implications of refinements
  in the input on $|J|$, $\delta$ and CKM matrix elements have also
  been studied.  
\end{abstract} 
 
 Recent discovery of neutrino oscillations in atmospheric neutrinos by
 the SuperKamiokande Collaboration \cite{superkam} has not only given
 the first clear cut signal for physics beyond the
 standard model (SM), but has also triggered great
 amount of activity in neutrino mixing phenomena as well as in
 the related issue of  fermion mass matrices. This has
 also  given an impetus to  study more deeply the
 quark mixing  phenomena which have been under investigation for the
 last two decades. In fact, there is a need to closely examine the
 quark mixing phenomena in the hope that one may decipher some signal,
 howsoever faint, for physics beyond the SM.

  In the context of quark mixing phenomena, over the last two
   decades, several analyses have been carried out
   \cite{review}, some of these
    in the last few years \cite{recent}.
   The basic purpose of these analyses has been the evaluation of
   Cabibbo, Kobayashi and Maskawa matrix ($V_{CKM}$) elements
    defined as, \\
   \be \left( \ba {c} d^{'} \\ s^{'}\\ b^{'} \ea \right) =
     \left( \ba {ccc} V_{ud} & V_{us} & V_{ub} \\
     V_{cd} & V_{cs} & V_{cb} \\ V_{td} & V_{ts} & V_{tb} \ea
      \right) \left( \ba {c} d \\ s\\ b \ea \right). \ee
      The usual inputs for the analyses are the CP violating
      parameters, $\epsilon_K$ and  $\epsilon_K^{'}$, as well
      as $B_o - \bar{B_o}$ mixing  phenomenon besides unitarity
      of CKM matrix defined as\\
    \be \sum_{\alpha=d,s,b} V_{i\alpha}V^*_{j\alpha}=\delta_{ij},
      \label{unit1} \ee
   \be \sum_{i=u,c,t} V_{i\alpha}V^*_{i\beta}=\delta_{\alpha
   \beta}.    \label{unit2} \ee
   where Latin subscripts run over the up type quarks $(u,c,t)$
   and Greek ones run over the down type quarks $(d,s,b)$.
   These analyses have given considerable insight into
    the dynamics of CKM matrix elements and their
  consequences, in particular the unitarity triangle
  (UT) based analyses \cite{hamz,ut,branco} have considerably sharpened the
   relationship between the CP violation and $B$ - decays. However,
   it is to be noted that in these analyses, the effect of unitarity,
   $\epsilon_K$ and $B_o - \bar{B_o}$ mixing etc. on the CKM matrix
   elements is carried out simultaneously. In other words,
   the separate implications of unitarity, $\epsilon_K$ and
   $B_o - \bar{B_o}$ mixing have not been studied, in particular,
   on such important quantities as CP violating phase $\delta$ and
   CKM matrix elements involving $t$ quark.

    In view of the availability of rephasing the quark fields \cite{review},
    the CKM matrix has 36 representations, therefore it has
    been advocated in the literature  that the analysis of
     CKM phenomenology should be carried out in a rephasing invariant
     manner \cite{hamz,rephase, jarlskog1}. In this context Jarlkog
     \cite{jarlskog1} has defined an
     interesting quantity $J$ which is rephasing invariant as well
     as all CP violating effects within the CKM paradigm are
     proportional to it. Interestingly $J$ is also directly related to
     the commutator of the quark mass matrices \cite{jarlskog1},
    for example,
       \begin{eqnarray}
       det[mm^{\dagger},m^{'}m^{'\dagger}] & = &
            -2iJ(m_t^2-m_c^2)(m_c^2-m_u^2)(m_u^2-m_t^2) \nonumber\\
           &  & \mbox{} \times(m_b^2-m_s^2)(m_s^2-m_d^2)
             (m_d^2-m_b^2).  \end{eqnarray}
     Therefore an evaluation of $|J|$ based
      on data is going to have important implications for
      texture specific mass matrices in the sense that it
     could provide valuable clues for searching the right texture
     for fermion mass matrices \cite{m.m}.

      In the very recent
     analyses, Parodi {\it et al} \cite{parodi} and Mele \cite{mele}
     primarily
     concentrate on finding CKM parameters in the Wolfenstein
 parametrization \cite{wolf} and the angles of unitarity triangle, while
     J. Swain {\it et al} \cite{swain} determine CKM parameters with
     and without unitarity.
      The PDG analysis \cite{pdg,pdg2k} evaluates the CKM matrix elements
      using the measured CKM elements and the unitarity as implied
   by the nine equations given by \ref{unit1} and \ref{unit2}.
    These analyses, however, have
   not been carried out in the rephasing invariant manner as well
  as they do not evaluate $J$.
  Therefore a rephasing invariant analysis of the CKM matrix
  based on unitarity is very much desirable in the hope that this
  may sharpen the predictions of unitarity for CKM matrix elements.

    The purpose of the present Rapid Communication is to evaluate
    $|J|$, based on non zero CP violation and on the unitarity triangle
    expressed by the relation,
    \begin{center}
  \be  V_{ud}V_{cd}^{*} + V_{us}V_{cs}^{*} + V_{ub}V_{cb}^{*} = 0
     \label{uc} \ee
  \end{center}
  and referred to as $uc$ triangle.
   This is the only unitarity triangle out of the six implied
   by equations \ref{unit1}
   and  \ref{unit2} with $i \neq j$ and $\alpha \neq \beta$, which
   involves well determined CKM matrix elements.
    After evaluating $|J|$, we use the PDG
    representations of $V_{CKM}$ to find CP violating phase
    $\delta$ and the elements of CKM matrix involving $t$
    quark. We also intend to examine the implications of
    present as well as future refinements
    in measured $V_{CKM}$ elements on fixing $\delta$.

  To begin with, we evaluate $|J|$, defined as \cite{jarlskog2} \\
   \be Im[V_{\alpha j}V_{\beta k}V_{\alpha k}^{*}V_{\beta j}^{*}] = J
  \sum_{\gamma,l}\epsilon_{\alpha,\beta,\gamma}\epsilon_{j,k,l}. 
      \label{j} \ee
   In principle one can evaluate $|J|$ using the above formula,
   however, in practice it does not help much as it involves CP
   violating phase $\delta$, the least known CKM parameter.
   Therefore,  for the purpose of our analysis, we exploited the
     relationship of $|J|$ with the unitarity triangle.
     Out of the
      six possible unitarity triangles
     we have used the  triangle expressed through the equation \ref{uc}.
 As mentioned earlier this triangle involves only those CKM
    elements which have been directly
   measured, consequently $|J|$ can be evaluated through
    the relation,\\
 \be  |J| = 2 \times Area ~~of~~ the~~ Unitarity ~~Triangle.
 \label{area} \ee
 With the availabilty of PDG2000 CKM elements \cite{pdg2k}, it is 
 natural to use 
 these as input  for evaluating $|J|$, however, with a view to
 understand the effect of refinements in CKM matrix elements on $|J|$,
 we have also done our calculations with PDG98 \cite{pdg} values. 
 In the same vein 
 we have also carried our calculations for ``future" values of CKM 
elements which may be available in future. In table
 \ref{tabinp} we have given the  PDG98 
 \cite{pdg} values of $V_{CKM}$ elements,
 \vud, \vus, \rub, \vcd, \vcs~ and \vcb,~  as well as 
  the recent PDG2000 values and ``future" values. 
 While listing the ``future" values, we have considered only those
elements in which the present error is more than $15\%$.
  
\begin{figure}
   \centerline{\psfig{figure=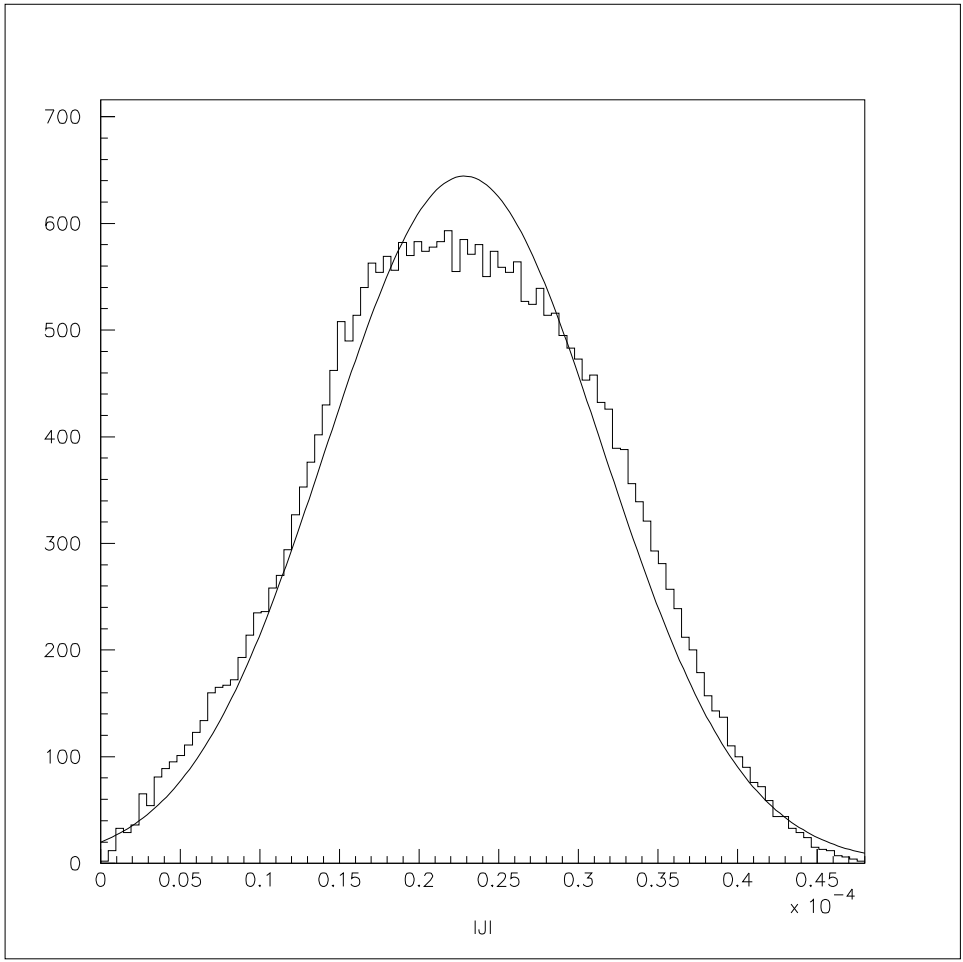,width=5in,height=5in}}
   \caption{The histogram as well as the gaussian fit of $|J|$ generated
   by considering input parameters at their  $90 \%$
  confidence level.}
  \label{fig1} 
  \end{figure}

 Before proceeding further, it is to be noted that the triangle
  mentioned above is highly squashed. The sides of the triangle 
  represented by $|V_{ud}^*V_{cd}|~(=a)$ and
  $|V_{us}^*V_{cs}| ~(=b)$ are of comparable lengths while the third side
  $|V_{ub}^*V_{cb}| ~(=c)$  is several orders of
    magnitude smaller compared to $a$ and $b$.
  This creates complications for evaluating the area of the triangle
  without violating unitarity and the existence of CP violation.
  To avoid these complications we have used the constraints
   $|a|+|c| > |b|$ and $|b|+|c| > |a|$ as suggested by Branco and Lavoura
 \cite{branco}.
   Using these
  constraints and the experimental data given in the column II of
     table \ref{tabinp} (PDG98), we have
  generated a histogram shown in figure \ref{fig1}.
  In generating the histogram  all inputs i.e. \vud,~\vus,~\rub,
  ~\vcd, ~\vcs ~and ~\vcb ~ have been taken 
     at their $90 \%$
  confidence level facilitating comparison with the corresponding PDG analysis.
   A Gaussian is fitted into the histogram plotted with approximately
 30,000 entries
   as shown in
  figure \ref{fig1}. The resulting value of  $|J|$
   is given as,
     \be |J|= (2.28 \pm 0.86) \times 10^{-5}  
          \label{jpdg1s},  \ee
  which in the $90\%$ C.L. leads to the range,
  \be |J|= (0.87 - 3.69) \times 10^{-5}.
          \label{jpdg90}  \ee
  The value of $|J|$
  can now be used to calculate $\delta$ using the PDG representations
  of CKM matrix, for example,
  \be V_{CKM}= \left( \ba {lll} c_{12} c_{13} & s_{12} s_{13} &
  s_{13}e^{-i\delta} \\
  -s_{12} c_{23} - c_{12} s_{23} s_{13}e^{i\delta} &
  c_{12} c_{23} - s_{12} s_{23}s_{13}e^{i\delta} 
  & s_{23} c_{13} \\
  s_{12} s_{23} - c_{12} c_{23} s_{13}e^{i\delta} &
  - c_{12} s_{23} - s_{12}c_{23} s_{13}e^{i\delta} &
  c_{23} c_{13} \ea \right),  \label{ckm} \ee
  with $c_{ij}=cos\theta_{ij}$ and   $s_{ij}=sin\theta_{ij}$ for the
  generation labels  $i,j=1,2,3.$
    In the above representation, $J$ can be expressed as,
      \be J = J^{'} sin \delta  \label{jpdg}, \ee where,  
        \be J^{'} = sin\theta_{12}sin\theta_{23}sin\theta_{13}cos\theta_{12}
        cos\theta_{23}cos^2\theta_{13}. \label{j'} \ee
 
   Before evaluating $J^{'}$, using equation \ref{tabinp} we calculate
    $ sin\theta_{12}, sin\theta_{23}$ and $ sin\theta_{13} $ from
    the experimental values of  \vus, \rub,~ 
    and  \vcb~ given in table \ref{tabinp}. 
    The corresponding values of  $ sin\theta_{12}, sin\theta_{23}$
 $ sin\theta_{13} $, presented in table \ref{tabsin}, are used
 to evaluate $J^{'}$. 
   Following the procedure outlined above for evaluating $|J|$, 
 with all input values at their $90\%$ C.L.,
     $J^{'}$ comes out to be,
      \be J^{'}= (2.86 \pm 0.76) \times 10^{-5}.
              \label{xpdg1s} \ee
    The range corresponding to $90\%$ C.L. of $J^{'}$ can be easily found
   out and is given as, 
     \be J^{'}=   (1.61 - 4.11) \times 10^{-5}. \label{xpdg90} \ee
 
\begin{figure}
   \centerline{\psfig{figure=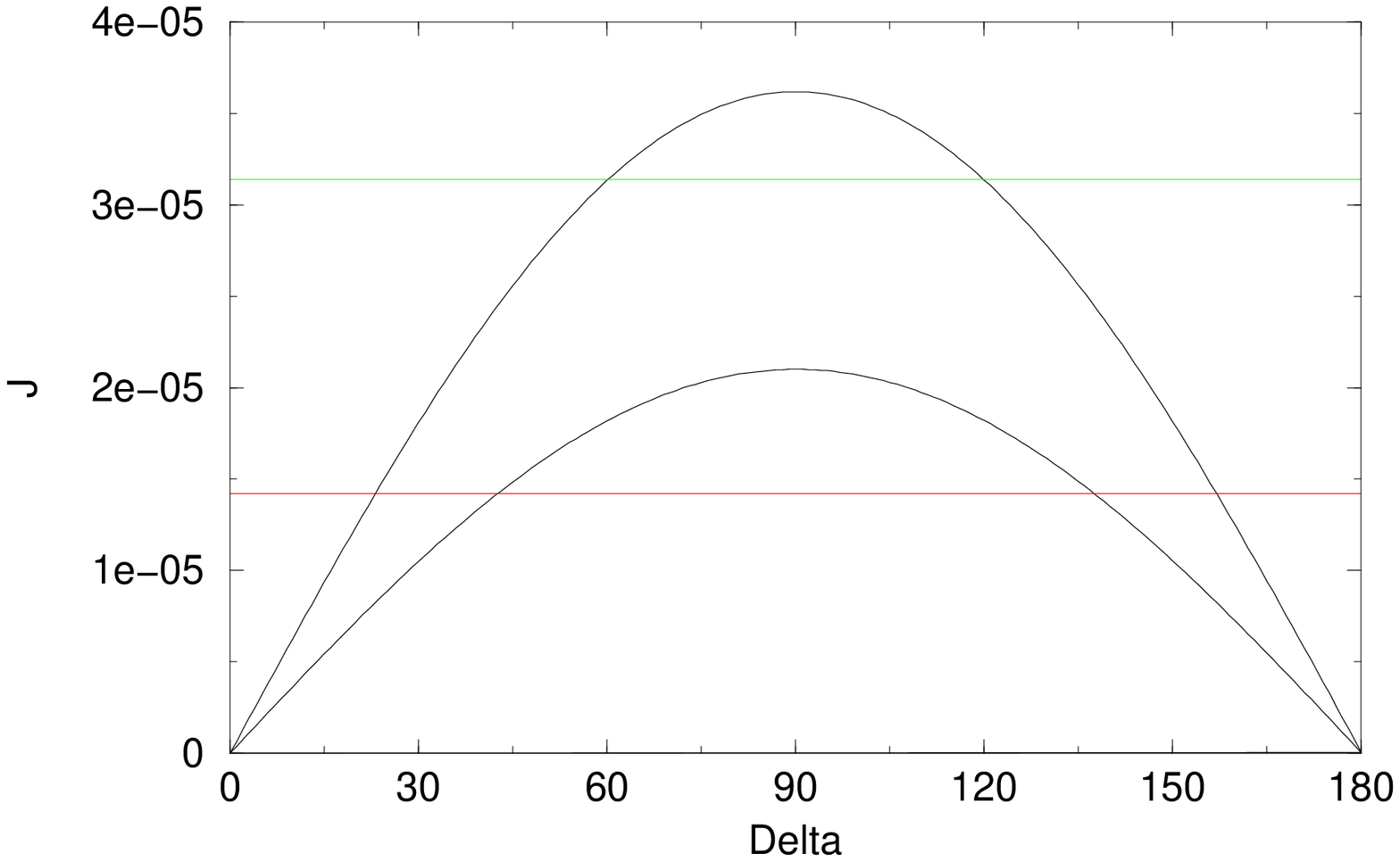,width=5in,height=5in}}
   \caption{Plot of $J(=J^{'}sin\delta)$ vs. $\delta$.  
  The upper and lower
 sinusoidal curves correspond to upper and lower limits of $J^{'}$ given
 by equation
 \ref{xpdg1s}.  The horizontal
 lines depict upper and lower limits of $|J|$ given by equation
  \ref{jpdg1s}.}
  \label{fig2}
   \end{figure}

\begin{figure}
   \centerline{\psfig{figure=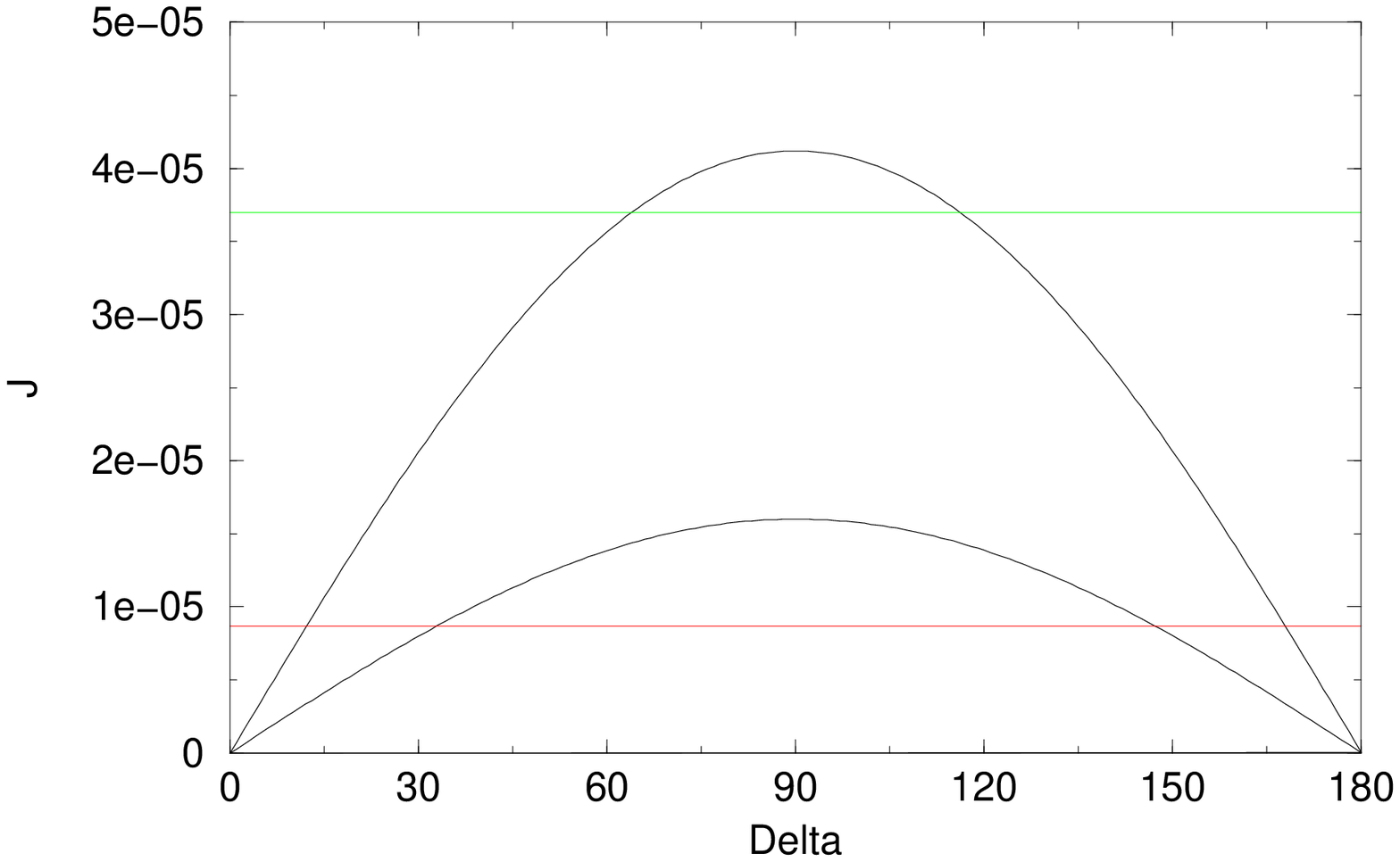,width=5in,height=5in}}
     \caption{Plot of $J(=J^{'}sin\delta)$ vs. $\delta$.  
  The upper and lower
 sinusoidal curves correspond to upper and lower limits of $J^{'}$ given
 by equation
 \ref{xpdg90}.  The horizontal
 lines depict upper and lower limits of $|J|$ given by equation
  \ref{jpdg90}.} 
   \label{fig3}
    \end{figure}

   Using equation \ref{jpdg}, we can find the range of $\delta$ corresponding
 to various C.Ls. of $|J|$ and $J^{'}$. 
   In this regard in figure \ref{fig2}, we have  
    plotted $J^{'} sin\delta$ as a function of $\delta$. The upper and lower
 sinusoidal curves correspond to upper and lower limits of $J^{'}$ given 
 by equation
 \ref{xpdg1s}.  The horizontal
 lines depict upper and lower limits of $|J|$ given by equation 
  \ref{jpdg1s}. Since $J^{'} sin \delta$ should reproduce  $|J|$ 
 calculated through the unitarity triangle $uc$, therefore from figure 
\ref{fig2}, by comparing the two  
    one can easily find out the widest limits on $\delta$, for example,\\
 \be    \delta = 23^o ~to  ~157^o. \label{del1s} \ee
       The above range of $\delta$ corresponds to $1\sigma$ C.L. of 
  both $|J|$ and $J^{'}$.
   Similarly, using \ref{jpdg90} and \ref{xpdg90}, the corresponding range of
  $\delta$  at $90\%$ C.L. of $|J|$ and $J^{'}$ can be found out as shown in
  figure \ref{fig3} and is given as,
  \be  \delta = 12^o ~to ~168^o. \label{del90} \ee

       These values of $\delta$ apparently look to be the consequence
  only of the unitarity relationship given by equation \ref{uc}.
    However on further investigation, as shown by Branco and Lavoura
  \cite{branco}, one finds that these $\delta$ ranges are consequences
   of all the non trivial unitarity constraints. In this sense the above
  range could be attributed to as a consequence of unitarity of the 
 CKM matrix.
   
     Alternatively, using equation \ref{jpdg}, one can find out
   $\delta$ for each of the 30,000 entries and likewise plot a
   histogram for $\delta$. Again fitting a Gaussian to the histogram,
    $\delta$ comes out to be,
 \beqn \delta & = & 51^o \pm 21^o~ (I~ quadrant), \nonumber \\
     & &  129^o \pm 21^o~ (II~ quadrant). \label{dhis68} \eeqn
 The corresponding range of $\delta$ at $90\%$ C.L. is,
     \beqn \delta & = & 17^o~to~  85^o~ (I~ quadrant), \nonumber \\
     & &  95^o~to~163^o~ (II~ quadrant). \label{dhis90} \eeqn
 This gives us relatively stronger bounds on $\delta$.  
 However, for the
 subsequent calculations we have used ranges of $\delta$ as given by
 \ref{del1s} and \ref{del90}.
 
    As has been mentioned earlier, we intend to compare our results
    both with PDG98 \cite{pdg} as well as with PDG2000 \cite{pdg2k}
    CKM matrix. To begin with, as the input for $|J|$ has been from PDG98
    values, therefore we compare our results with PDG98 CKM matrix.
     To calculate the CKM elements, we have  used the 
     values of sines of mixing angles at their $90\%$ C.L. (using column
   II of table \ref{tabsin}) and
     $\delta$ at $90\%$ C.L. of $|J|$ (equation \ref{del90}).
   The calculated
      $V_{CKM}$ matrix is,
 \be   \left[ \ba {ccc} 0.9747-0.9765 &   0.216-0.223 & 0.0019-0.0045 \\
    0.216-0.223 & 0.9739-0.9757 & 0.037-0.042 \\
    0.004-0.014 & 0.035-0.042 & 0.9992 \ea \right].
 \label{ckm98cal} \ee
  To facilitate the comparison of corresponding CKM matrix elements as
  well as for easy readability, we present below the $V_{CKM}$
  from the PDG98 \cite{pdg} also calculated at $90\%$ C.L..
   \be   \left[ \ba {ccc}  0.9745-0.9760 & 0.217-0.224 & 0.0018-0.0045 \\
    0.217-0.224 & 0.9737-0.9753 &  0.036-0.042 
   \\  0.004-0.013 & 0.035-0.042 & 0.9991-0.9994 \ea \right].
  \label{ckmpdg98} \ee
    Comparing \ref{ckm98cal}  and \ref{ckmpdg98}, one finds
    that we have been able to reproduce PDG matrix with minor differences 
   only at fourth decimal places. In addition to the reproduction of CKM
     matrix elements, it needs to be emphasized that we have
     calculated $|J|$ and $\delta$ based entirely on unitarity and data,
     which to our knowledge has not been calculated earlier. The
   availabilty of $|J|$ and $\delta$ simplifies the task of calculating
   CKM matrix elements as well as giving a deeper insight into the
   contribution of CKM paradigm to CP violating phenomena.
  
   The above method of evaluating  $|J|$, $\delta$ and CKM matrix
    elements  can be easily repeated for the PDG2000 as well as
 ``future" values of input parameters mentioned in tables \ref{tabinp}
 and \ref{tabsin}. The corresponding $|J|$ and $\delta$ have been
 given in table \ref{tabj}. 
   For the sake of completeness, we have repeated the whole analysis
  with input values at their $1 \sigma$ and $3 \sigma$ C.Ls. and
  the corresponding results for $|J|$, $J^{'}$ and $\delta$
   are listed in table \ref{tabj}.
 
A close look at the table \ref{tabj} leads us to several interesting
  points. For example, the bounds on $|J|$ and $\delta$ are weak when the 
  input values are taken at their $3 \sigma$ C.L., but we get relatively
  stronger bounds when the analysis is done with the input values being
  at their $90\%$ and $1 \sigma$ C.L.. Further, it is clear from
  the same table that there is not much change in the range of $\delta$
  with PDG98 and PDG2000 values. However, if in future the ranges of
   \vcs~ and
  \rub ~ are further constrained, lt may lead to significant narrowing in
  the range of $\delta$ so obtained, as shown in columns V and VI of
  row IV of table \ref{tabj}.

  To give a better insight into the implications of PDG2000 values on
  our calculations, we present below the CKM matrix evaluated using
    values of sines of mixing angles at their $90\%$ C.L. (from column 
  III of table \ref{tabsin}) and 
     $\delta$ at $90\%$ C.L. of $|J|$.
  
    The calculated  $V_{CKM}$ matrix is,
 \be   \left[ \ba {ccc} 0.9756-0.9765 &   0.216-0.223 & 0.002-0.005 \\
    0.216-0.223 & 0.9739-0.9757 & 0.037-0.043 \\
    0.005-0.013 & 0.036-0.043 & 0.9992 \ea \right].
    \label{ckm2kcal}\ee
  To facilitate the comparison, we present below the $V_{CKM}$
  from the PDG2000 \cite{pdg2k} also calculated at $90\%$ C.L..
   \be   \left[ \ba {ccc}  0.9742-0.9757 & 0.219-0.226 & 0.002-0.005 \\
    0.219-0.225 & 0.9734-0.9749 &  0.037-0.043 \\
    0.004-0.014 &  0.035-0.043 & 0.9990-0.9993 \ea \right].
     \label{ckm2kpdg} \ee
      Comparing the CKM matrices \ref{ckm2kcal} and \ref{ckm2kpdg},
  we see that we are able to reproduce PDG2000 CKM matrix,
  once again justifying the procedure followed in carrying out the
  present analysis.
  To assess the impact of future refinements in CKM matrix elements
  \rub~ and \vcs, ~we present below the CKM matrix elements corresponding
  to $|J|$ and $\delta$ as mentioned in row III of table \ref{tabj}.
 \be   \left[ \ba {ccc} 0.9747-0.9765 &   0.216-0.223 & 0.0029-0.0043 \\
    0.216-0.223 & 0.9739-0.9757 & 0.037-0.043 \\
    0.006-0.013 & 0.036-0.043 & 0.9992 \ea \right].
    \label{ckmfutcal}\ee
  Comparing \ref{ckm98cal}, \ref{ckm2kcal} and \ref{ckmfutcal}
        we find that the $|V_{CKM}|$ matrix elements
       do not show much variation when the
      latest \cite{pdg2k} or the ``future" values are used.
      The small changes in the CKM elements mostly at the fourth 
      decimal places  are
      not primarily due to change in the range of $\delta$,  but
      due to overall changes in all the input parameters.
      This probably restricts the use of unitarity in evaluating
      the  $|V_{CKM}|$ elements involving $t$ quark.

     In conclusion, we would like to mention that using 
     only one of the six unitarity triangles, we
     have evaluated Jarlskog rephasing invariant parameter $|J|$
      and consequently $\delta$. Using this range
     of $\delta$ we have been able to reproduce the PDG matrix at
     $90\%$  C.L. evaluated by PDG group.
     Our calculations also indicate that improvements and further
      refinements in $V_{CKM}$ elements \vcs~ and \rub~ result in
      significant narrowing in the range of $\delta$, however there is
      no appreciable impact on $V_{CKM}$ elements involving $t$ quark,
       therefore, their range can be narrowed only by direct
       measurement of $\delta$.
      It needs to be mentioned that an evaluation of $|J|$ based
      on data is going to have important implications for
      texture specific mass matrices as the parameter $J$ is
       directly related to the mass matrices.
    Our conclusions in this regard would be published elsewhere.

\vskip 1cm
  {\bf ACKNOWLEDGMENTS}\\
 M.R. would like to thank CSIR, Govt. of India, for
 financial support.
M.R. and P.S.G. would like to thank the chairman, Department of Physics,
for providing facilities to work in the department. P.S.G. acknowledges
the financial support received for his UGC project.

 \pagebreak
  \begin{table}
  \begin{tabular}{|l|l|l|l|}   \hline
  Element & PDG98 values  & PDG2000 values  & ``Future"
  values\\ \hline
  \vud & 0.9740 $\pm$ 0.0010 &  0.9735 $\pm$ 0.0008 &  0.9735 $\pm$ 0.0008\\
  \vus & 0.2196 $\pm$ 0.0023 & 0.2196 $\pm$ 0.0023 & 0.2196 $\pm$ 0.0023\\
  \rub & 0.08 $\pm$ 0.02 & 0.090 $\pm$ 0.025  & 0.090 $\pm$ 0.010\\
  \vcd & 0.224 $\pm$ 0.016 & 0.224 $\pm$ 0.016 & 0.224 $\pm$ 0.016\\
  \vcs & 1.04 $\pm$ 0.16 & 1.04 $\pm$ 0.16 & 1.04 $\pm$ 0.08\\
  \vcb & 0.0395 $\pm$ 0.0017 & 0.0402 $\pm$ 0.0019
  & 0.0402 $\pm$ 0.0019 \\ \hline
  \end{tabular}
  \caption{The experimental values of $V_{CKM}$ elements, \vud,
  \vus,~\rub, \vcd, \vcs~ and  \vcb.}
  \label{tabinp} \end{table}

% \pagebreak
    \begin{table}
    \begin{tabular}{|l|l|l|l|}   \hline
    Parameter & Input PDG98 values   &  Input PDG2000 values &
    Input ``Future" values \\ \hline
    $ sin\theta_{12}$ & $0.2196 \pm 0.0023$ & $0.2196 \pm 0.0023$
      & $0.2196 \pm 0.0023 $ \\
    $ sin\theta_{23}$ & $0.0395 \pm 0.0017$ & $0.0402 \pm 0.0019$
     & $0.0402 \pm 0.0019$ \\
     $ sin\theta_{13}$ & $0.0032 \pm 0.0008$ & $0.0036 \pm 0.0010$
     & $ 0.0036\pm 0.0004 $\\  \hline
    \end{tabular}
    \caption{Sines of the mixing angles calculated from the data
     given in table \ref{tabinp}.}
    \label{tabsin} \end{table}

 %   \pagebreak
{\small
   \begin{table}
    \begin{tabular}{|l|l|l|l|l|l|}   \hline
& &  $~~~~~~|J|$  &  $~~~~~~~~J^{'}$  &
\bt{l} $\delta$ obtained\\ graphically  with\\ $|J|$ and $J^{'}$ \\
    at $1 \sigma$  C.L. \\ \et &
\bt{l} $\delta$ obtained\\ graphically  with\\ $|J|$ and $J^{'}$ \\
    at $90\%$  C.L. \\ \et \\ \hline
% -----------------------------------------------------
   &   \bt{l} All inputs\\ at $1\sigma$ C.L.\\ \et &
  \bt{l} $(2.21 \pm .62)$ \\ $\times 10^{-5}$ \\ \et
 &   \bt{l}  $(2.76 \pm .44)$ \\ $\times 10^{-5}$ \\ \et
 &  $ 30^o - 150^o$ & $ 20^o - 160^o$ \\
& & & & & \\
    PDG98 & \bt{l} All inputs \\  at $90\%$ C.L. \\ \et &
  \bt{l} $(2.28 \pm .86)$ \\ $\times 10^{-5}$ \\ \et
&  \bt{l} $(2.86 \pm .76)$ \\ $\times 10^{-5}$ \\ \et
 & $23^o - 157^o$  &  $12^o - 168^o$ \\
& & & & & \\
& \bt{l} All inputs \\ at $3\sigma$ C.L. \\ \et &
     \bt{l} $(2.65 \pm 1.37)$ \\ $\times 10^{-5}$ \\ \et
&  \bt{l} $(3.26 \pm 1.30)$ \\ $\times 10^{-5}$ \\ \et
 & $16^o - 164^o$   &  $4^o - 176^o$ \\ \hline
% ----------------------------------------------------
 &   \bt{l} All inputs\\ at $1\sigma$ C.L.\\ \et &
  \bt{l} $(2.59 \pm .79)$ \\ $\times 10^{-5}$ \\ \et
 &   \bt{l}  $(3.23 \pm 0.63)$ \\ $\times 10^{-5}$ \\ \et
 &  $ 28^o - 152^o$ & $ 18^o - 162^o$ \\
& & & & & \\
    PDG2000 & \bt{l} All inputs \\  at $90\%$ C.L. \\ \et &
  \bt{l} $(2.71 \pm 1.12)$ \\ $\times 10^{-5}$ \\ \et
&  \bt{l} $(3.41 \pm 1.06)$ \\ $\times 10^{-5}$ \\ \et
 & $21^o - 159^o$  &  $10^o - 170^o$ \\
& & & & & \\
& \bt{l} All inputs \\ at $3\sigma$ C.L. \\ \et &
     \bt{l} $(3.29 \pm 1.78)$ \\ $\times 10^{-5}$ \\ \et
&  \bt{l} $(4.14 \pm 1.67)$ \\ $\times 10^{-5}$ \\ \et
 & $15^o - 165^o$   &  $3^o - 177^o$ \\ \hline
% -----------------------------------------------------
 &   \bt{l} All inputs\\ at $1\sigma$ C.L.\\ \et &
  \bt{l} $(2.79 \pm 0.49)$ \\ $\times 10^{-5}$ \\ \et
 &   \bt{l}  $(3.14 \pm 0.31)$ \\ $\times 10^{-5}$ \\ \et
 &  $ 42^o - 138^o$ & $ 32^o - 148^o$ \\
& & & & & \\
   \bt{l} ``Future" \\ values \\ \et
   & \bt{l} All inputs \\  at $90\%$ C.L. \\ \et &
  \bt{l} $(2.61 \pm 0.78)$ \\ $\times 10^{-5}$ \\ \et
&  \bt{l} $(3.19 \pm 0.49)$ \\ $\times 10^{-5}$ \\ \et
 & $30^o - 150^o$  &  $19^o - 161^o$ \\
& & & & & \\
& \bt{l} All inputs \\ at $3\sigma$ C.L. \\ \et &
     \bt{l} $(2.73 \pm 1.09)$ \\ $\times 10^{-5}$ \\ \et
&  \bt{l} $(3.34 \pm 0.89)$ \\ $\times 10^{-5}$ \\ \et
 & $23^o - 157^o$   &  $11^o - 169^o$ \\ \hline
 \end{tabular}
    \caption{ $|J|$,  $J^{'}$ and $\delta$ values  obtained
  by using PDG98, PDG2000 and ``future values" of input parameters
  listed in table \ref{tabinp} and \ref{tabsin} at their
  $1\sigma$, $90\%$ and $3 \sigma$ C.L.'s}
    \label{tabj} \end{table}}

\newpage

\begin{thebibliography}{99}
 
\bibitem{superkam} Superkamiokande collaboration; Y. Fukuda et. al.,
   Phys. Lett. {\bf B433}, 9(1998) and {\bf B436}, 33(1998);\\
  T Kajita, Talk presented at Neutrino-98, Takayama, Japan (1998).
   
 \bibitem{review} CP violation, Ed. L. wolfenstein, North Holland,
 elsevier Science Publishers B.V., 1989;
 CP violation, Ed. C. Jarlskog, World Scientific Publishing Co. Pte.
  Ltd, 1989;
  G. Buchalla, Andrzej J. Buras, M. E. Lautenbacher, Rev. Mod. Phys.
   {\bf 68}, 1125(1996) and references therein;

  \bibitem{recent}
   Manmohan Gupta and P. S. Gill, Pramana {\bf 38}, 477(1992); 
  Stefan Herrlich and Ulrich Nierste, Phys. Rev. {\bf D52},
 6505(1995);
 M. Gronau and J. L. Rosner, Phys. Rev. Lett. {\bf 76}, 1200(1996);
 A. Buras, in ``Probing the Standard Model of Particle Interactions",
 F. David and R. Gupta, Eds, 1998 Elsevier Science B. V.;
   P. S. Gill and Manmohan Gupta, Mod. Phys.
 Lett. {\bf A13}, 2445(1998).
 H. Fritzsch and Z. Z. Xing, Nucl. Phys. {\bf B556}, 49(1999);
 A. Ali, in Proc. of the 13th Topical Conference on Hadron Collider
  Physics, TIFR, Mumbai, India (1999);
  I. I. Bigi and A. I. Sanda, hep-ph/9909479.

 \bibitem{hamz} A. Campa, C. Hamzaoui and V. Rahal, Phys. Rev. {\bf D39},
 3435(1989).

 \bibitem{ut}
  Stefan Herrlich and Ulrich Nierste, Phys. Rev. {\bf D52},
 6505(1995);
 M. Gronau and J. L. Rosner, Phys. Rev. Lett. {\bf 76}, 1200(1996);
 H. Fritzsch and Z. Z. Xing, Nucl. Phys. {\bf B556}, 49(1999);
 A. Ali, in Proc. of the 13th Topical Conference on Hadron Collider
  Physics, TIFR, Mumbai, India (1999);
  I. I. Bigi and A. I. Sanda, hep-ph/9909479.

\bibitem{branco} G.C. Branco and L. Lavoura, Phys. Lett. {\bf B208},
 123(1988).

\bibitem{rephase} Y. Koide et al., Univ Shizuoka, US91-04(1991);
 G. Belanger, E. Boridy, C. Hamzaoui and G. G. Jakimov,
 Phys. Rev. {\bf D48}, 4275(1993).

 \bibitem{jarlskog1} C. Jarlskog, Phys. Rev. Lett. {\bf 55},
 1039(1985); Zeit f. Phys. {\bf C29}, 491(1985).

 \bibitem{m.m} A. Campa, C. Hamzaoui and V. Rahal, Phys. Rev. {\bf D39},
 3435(1989); P. S. Gill and Manmohan Gupta, J. Phys {\bf G23},
 335(1997); Phys. Rev. {\bf D56} 3143(1997); H. Fritzsch and
 Z. Z. Xing, Nuc. Phys. {\bf B556}, 49(1999). 

 \bibitem{parodi} F. Parodi, P. Roudeau and A. Stocchi, Nuovo Cim.
   {\bf A112}, 833(1999).

  \bibitem{mele} S. Mele, hep-ph/9808411, Proceedings of workshop on CP
  violation, Adelaide, Australia, July 3-8(1998).

  \bibitem{wolf} L. Wolfenstein, Phys. Rev. Lett. {\bf 51}, 1945(1983).

 \bibitem{swain} John Swain and Lucas Taylor, hep-ph/9712421.

 \bibitem{pdg} C. Caso et. al., Particle Data Group, Euro. Phys. J.
 {\bf C3}, 1(1998).

 \bibitem{pdg2k} D.E. Groom et. al., Particle Data group, Euro. Phys. 
 J. {\bf C15}, 1(2000). 

\bibitem{jarlskog2} C. Jarlskog, Proc. of 1986 Rencontre de moriond
 on ``Progress in electroweak interactions", (Ed. J. Tran Thanh
 Vass) p. 389.

%  \bibitem{rub} F. Parodi, in Proc. of XXIXth Int. Conf. on High
 % Energy Physics, Vancouver, B.C., (1998).
  %CLEO Collab., A. Bean {\it et al.} Phys. Rev. Lett.
 % {\bf 70}, 2861(1993); CLEO Collab., J.P. Alexander {\it et al.},
 %  Phys. Rev. Lett. {\bf 77}, 5000(1996).

 %\bibitem{wolf1} M. Kobayashi, Prog. Theor. Phys. {\bf 92},
 %287(1994); 289(1994); Z. Z. Xing, Phys. Rev. {\bf D51},
 %3958(1995).

\end{thebibliography}
\end{document}